\documentstyle[aps,pre,eqsecnum]{revtex} 
\input amssym.def \input amssym.tex 

\begin{document}
\draft

\title{The Exact Ground State of the Frenkel-Kontorova Model
with Repeated Parabolic Potential \\
 I. Basic Results 
 }

\author{R. B. Griffiths}  
\address{Department of Physics \\
         Carnegie-Mellon University\\
         Pittsburgh, PA 15213, USA}

\author{H. J. Schellnhuber}
\address{Potsdam Institute for Climate Impact Research (PIK)\\
P.O. Box 60 12 03
D-14412 Potsdam, Germany}

\author{H. Urbschat}
\address{Fachbereich Physik and ICBM \\
         Carl von Ossietzky Universit\"at Oldenburg\\
         Postbox 2503 \\
         D-26111 Oldenburg,Germany}

\date{Version of June 24, 1997}
\maketitle

\begin{abstract}
	The problem of finding the exact energies and configurations for the
Frenkel-Kontorova model consisting of particles in one dimension connected to
their nearest-neighbors by springs and placed in a periodic potential
consisting of segments from parabolas of identical (positive) curvature but
arbitrary height and spacing, is reduced to that of minimizing a certain convex
function defined on a finite simplex.
\end{abstract}

\section{Introduction}
\label{sec1}
	The Frenkel-Kontorova model of classical particles connected to their
nearest neighbors by springs and placed in a one-dimensional periodic potential
$V$ \cite{1,2,2a,2b} is of interest for several different reasons. First, it
provides a simple example which exhibits some of the complex phenomena
(discommensurations, devil's staircase phase diagrams) associated with
commensurate-incommensurate phase transitions \cite{3}.  For some recent
studies, see \cite{bg94,hls94}; also \cite{bbc94} for extensions to quantum
systems, and \cite{mk95} for classical particles at a finite temperature.
Next, various interesting dynamical phenomena result if forces, either static
or dynamic, are applied to the particles in a Frenkel-Kontorova chain; for
some recent work, see \cite{ff,st95}.  Finally, the equilibrium configurations
of this model form the orbits of an area-preserving twist map \cite{4}, so that
their study is part of the subject of nonlinear dynamical systems \cite{5}.

	A special, exactly solvable realization introduced by Aubry \cite{6}
and Percival \cite{7}, in which $V$ consists of a potential made up
of identical parabolas, has played a pioneering role in the study of general
Frenkel-Kontorova models. It is particularly useful for understanding the
properties of incommensurate ground states in the ``Cantorus" regime: such
configurations define invariant Cantor sets of the associated area-preserving
map, on which the motion is quasiperiodic with an irrational winding number
$w$.
Cantori, or Aubry-Mather sets \cite{8}, have been the subject of considerable
research \cite{9,9a,10,10a,11,12}, as they constitute the ubiquitous
remnants of
broken KAM tori in non-integrable Hamiltonian systems \cite{13}.

	The present paper is dedicated to a generalization of this solvable
model to the case in which the parabolas all have the same curvature, but need
not be uniformly spaced nor located at the same height.  The only constraint is
that after $N$ parabolas, where $N$ is finite, the whole pattern is repeated,
thus giving rise to a periodic $V$. See the example with $N=3$ sketched in
Fig.~\ref{fig3}. The case $N=1$ is, of course, the one studied earlier by Aubry
and Percival.

The generalization is important for two reasons: First, as we pointed out for
$N=2$ in a previous paper \cite{14}, qualitatively novel phenomena emerge which
are not present in the case $N=1$. Examples are incommensurate defects,
discontinuous Cantorus-Cantorus phase transitions, and independent orbits of
gaps composing the complement of the Aubry-Mather set.  Second, it provides a
tool, through the possibility of approximating a smooth potential by a set of
parabolas, for identifying true ground states in the presence of infinite sets
of competing metastable Cantorus configurations, as encountered in
Frenkel-Kontorova models with smooth $V$ differing from a simple cosine
\cite{15}.  

	In the following sections we give the detailed solution for general
$N$, which in essence consists in replacing the original infinite-variable
minimization problem by that of finding the minimum of a certain convex
function defined on a finite simplex. While the second minimization problem is
not altogether trivial, it can be carried out to quite adequate precision in
particular cases by a combination of analytical and numerical techniques.
Efficient numerical methods are the subject of the paper which follows.

	Previous research on this topic of which we are aware includes an
unpublished thesis of F. Vallet \cite{16}. He considered some particular cases
for $N>1$, including the $N=2$ situation with equally spaced parabolas
discussed in our previous publication, for which he wrote down an explicit
expression for the ground state energy. And he pointed out that the same
methods could be applied to the general $N$ situation using a parametrization
of the hull function very similar to the one employed here.  More recently Kao,
Lee, and Tzeng \cite{klt96b} have addressed the problem of finding the ground
state by using a hull function to find solutions of the equilibrium equations.
Their results are consistent with those in the present manuscript.

	In addition, Aubry, Axel and Vallet considered a somewhat different
extension of the $N=1$ solvable model, with even and odd numbered particles
subjected to different potentials \cite{17}.  While there are certain
similarities, this model (and the method used to solve it) seems to be distinct
from the $N=2$ case considered here.  Recently Kao, Lee, and Tzeng
\cite{klt96a}
have solved a model in which the parabolas have negative curvature; the results
are very different from the case of positive curvature discussed here.

Our material is organized as follows. Section~\ref{sec2} contains basic
definitions for the model including the explicit form of the potential $V$.  In
Sec.~\ref{sec3} the energy for an equilibrium state is rewritten with the help
of a Green's function in a form which is combined with some of Aubry's exact
results for ground states in Sec.~\ref{sec4}. The result is used in
Sec.~\ref{sec5} to set up an explicit minimization problem of a convex function
over a finite simplex, the solution to which yields both the energy and
configuration of the ground state.  The conclusion, Sec.~\ref{concl}, indicates
a number of possible generalizations.  Minor technical issues are discussed in
two appendices.

\section{Energy}
\label{sec2}  
We shall consider a system described by an energy
\begin{equation}
H = \sum_{j=-\infty}^{\infty} [\frac{1}{2} (u_{j+1}-u_{j})^2 + V(u_{j})],
\label{2.1}
\end{equation}
where $u_{j}$ is the position of the $j$-th particle, while
\begin{equation}
V(u) = V(u+P) = \min_{n} p(n;u)
\label{2.2}
\end{equation}
is a periodic potential made up of segments of parabolas, as illustrated by
Fig.~\ref{fig3}.  The $n$-th parabola is given by the formula
\begin{equation}
p(n;u) = {\textstyle {1\over 2}} \kappa (u-t_{n})^2 + h_{n}.
\label{2.3}
\end{equation}
We assume that the positions of the minima of these parabolas form a monotonic
periodic sequence,
\begin{equation}
t_{n} < t_{n+1} \ ; \quad t_{n+N} = P + t_{n}
\label{2.4}
\end{equation}
and the values at the minima are periodic,
\begin{equation}
h_{n+N} = h_{n},
\label{2.5}
\end{equation}
but otherwise arbitrary.
    
Because all the parabolas have the same curvature, the $u$ axis consists of
segments separated by a sequence of {\it {dividing points}}, (see
Fig.~\ref{fig3}),
\begin{equation}
\bar{t}_{n} \leq \bar{t}_{n+1}; \quad \bar{t}_{n+N} = P + \bar{t}_{n},
\label{2.6}
\end{equation}
with the property that for $u$ in the open interval ($\bar{t}_{m-1},
\bar{t}_{m}$) (which may be empty), the minimum in (\ref{2.2}) is achieved for
$n=m$, and for no other value of $n$. Thus this interval is uniquely associated
with the $m$'th parabola, and if $u_{j}$ falls in this interval, we shall say
that particle $j$ is ``in" parabola $m$, or $n(j)=m$. Given the periodicity
(\ref{2.6}), it is clear that two such intervals whose subscripts differ by $N$
will be of the same length, and will have corresponding values of $V(u)$. Thus
we shall say that such an interval, or the corresponding parabola, is of ``type
$l$", $1 \leq l \leq N$, if $m=l(\hbox{mod } N)$.

Let
\begin{equation}
\tau _{j} = t_{n(j)}
\label{2.7}
\end{equation}
be the minimum of the parabola containing the $j$-th particle, in the sense
just discussed. Then (\ref{2.1}) can be rewritten as
\begin{equation}
H = H_{0} + H_{1},
\label{2.8}
\end{equation}
where
\begin{equation}
H_{0} = \frac{1}{2} \sum_{j=-\infty}^{\infty} [(u_{j+1} - u_{j})^2 + \kappa
(u_{j} - \tau_{j})^2]
\label{2.9}
\end{equation} 
and
\begin{equation}
H_{1} = \sum_{j=-\infty}^{\infty} h_{n(j)}.
\label{2.10}
\end{equation}

For later analysis it is convenient to rewrite (\ref{2.9}) in terms of
\begin{equation}
\hat{u}_{j} = u_{j} - \tau_{j},
\label{2.11}
\end{equation}
the distance of the $j$-th particle from the minimum of the parabola which
contains it. One obtains
\begin{equation}
H_{0} = \frac{1}{2} \sum_{j=-\infty}^{\infty} \sum_{k=-\infty}^{\infty}
H_{a}(j-k) \hat{u}_{j} \hat{u}_{k} - \sum_{j=-\infty}^{\infty} \hat{u}_{j}
\Delta^2 \tau_{j} + \frac{1}{2}  
\sum_{j=-\infty}^{\infty} (\Delta \tau_{j})^2 ,    
\label{2.12}
\end{equation}
where
\begin{equation}
\Delta \tau_{j} = \tau_{j+1} - \tau_{j},\quad \Delta^2 \tau_{j} = 
\tau_{j+1} -2\tau_{j} +\tau_{j-1} =
\Delta \tau_{j} - \Delta \tau_{j-1}
\label{2.13}
\end{equation}
and
\begin{equation}
H_{a}(0)=2+ \kappa , \quad H_{a}(1) = H_{a}(-1) = -1, 
\quad H_{a}(k)=0 \hbox{ for } |k|>1.
\label{2.14}
\end{equation}

\section{Equilibrium States}
\label{sec3}

	An equilibrium state is one in which the force is zero on every
particle, which means $\partial H_{0} / \partial \hat{u}_{j} = 0$ for every
$j$, and thus, by (\ref{2.12}), 
\begin{equation}
\sum_{k=-\infty}^{\infty} H_{a}(j-k) \hat{u}_{k} = \Delta ^2 \tau_{j}
\label{5.1}
\end{equation}
If we assume that the $\Delta^2\tau_j$ are known and are bounded as a function
of $j$, the solution of (\ref{5.1}) may be written as
\begin{equation}
\hat{u}_{j} = \sum_{k=-\infty}^{\infty} G(j-k) \Delta ^2 \tau_{k}
\label{5.2}
\end{equation}
using the Green's function (decreasing at infinity)
\begin{equation}
G(k) = e^{-r|k|}/2 \sinh r ,
\label{5.3}
\end{equation}
which satisfies the equation
\begin{equation}
\sum_{k=-\infty}^{\infty} H_{a}(j-k) G(k-l) = \delta_{jl}.
\label{5.4}
\end{equation}
Here $r$ is a positive number defined by:
\begin{equation}
\kappa = (2 \sinh r/2)^2.
\label{5.5}
\end{equation}

	Multiplying (\ref{5.1}) by $\hat{u}_{j}$ and summing yields, see
(\ref{2.12}) and (\ref{5.2}):
\begin{equation}
H_{0} - \frac{1}{2} \sum_{j=-\infty}^{\infty} (\Delta \tau_{j}) ^2 =
-\frac{1}{2} \sum_{j=-\infty}^{\infty} \hat{u}_{j} \Delta ^2 \tau_{j} = -
\frac{1}{2} \sum_{j=-\infty}^{\infty} \sum_{k=-\infty}^{\infty} G(j-k) \Delta
^2 \tau_{j} \Delta ^2 \tau_{k}.
\label{5.6}
\end{equation}
With the help of the function
\begin{equation}
g(j) = G(j+1) + G(j-1) - 2G(j) + \delta_{j0} = (\tanh r/2) e^{-r|j|}.
\label{5.9}
\end{equation}
one can rewrite (\ref{5.2}) and (\ref{5.6}) as:
\begin{equation}
u_{j} = \sum_{k=-\infty}^{\infty} g(j-k) \tau_{k},
\label{5.7}
\end{equation}
\begin{equation}
H_{0} = \frac{1}{2} \sum_{j=-\infty}^{\infty} \sum_{k=-\infty}^{\infty}
g(j-k)\Delta \tau_{j} \Delta \tau_{k}.
\label{5.8}
\end{equation}

The fact that
\begin{equation}
\sum_{j=-\infty}^{\infty} g(j) = 1
\label{5.10}
\end{equation}
means that (\ref{5.7}) expresses the position of particle $j$ as a weighted
average of the positions of the parabolas occupied by the different particles.

	It is useful for later purposes to extend $g(j)$ to a function
$g(x)$ defined on the entire real axis by means of linear interpolation between
successive integers, (see Fig.~\ref{fig5}), which is to say:
\begin{equation}
g(x) := [1- \hbox{frac} (x)]g(\hbox{Int}(x)) + \hbox{frac} (x)g(1+\hbox{Int}
(x)) ,
\label{5.11}
\end{equation}
where $\hbox{Int}(x)$ is the largest integer not greater than $x$, and
$\hbox{frac}(x)$
stands for $x-\hbox{Int}(x)$. Note that the resulting function is convex for
$x>0$,
and also for $x<0$, and possesses a piecewise constant derivative with
discontinuities whenever $x$ is an integer.

	If the equilibrium configuration is monotonic in the sense that for all
j, 
\begin{equation}
	n(j+1) \geq n(j),
\label{}
\end{equation}
that is, $\Delta\tau_j\geq 0$, then one can write $H_{0}$,
(\ref{5.8}) in an alternative form with the help of the function,
\begin{equation}
I_{m} = \max \{ j \epsilon {\Bbb Z}\ |\ n(j) \leq m \},
\label{5.12}
\end{equation}
the number of the last particle in parabola $m$, or of the last particle
preceding this parabola if it is empty.  (One can think of $I_m$ as a sort of
inverse of the function $n(j)$.) If
\begin{equation}
\Delta t_{m} = t_{m+1} - t_{m}
\label{5.13}
\end{equation}
is the distance between the minima of two successive parabolas, it is easy to
show that
\begin{equation}
 \tau_{j+1} - \tau_{j} = \Delta \tau_{j} = \sum_{m=-\infty}^{\infty} (\Delta
t_{m}) \delta_{j,I_{m}}.
\label{5.14}
\end{equation}
Thus if particle $j$ is not the last particle in some parabola, which is to say
particle $j+1$ is in the same parabola, $I_{m}$ never takes the value $j$ and
$\Delta \tau_{j} = 0$, so both sides of this equation vanish. If $j$ is the
last particle in parabola $m$ and parabola $m+1$ is not empty, then $\Delta
\tau_{j} = \Delta t_{m}$, $I_{m} = j$, and $I_{m+1}$ is larger than $j$, so
again (\ref{5.14}) holds. If, on the other hand, parabola $m$ is followed by
one or more empty parabolas, then for each of these $I$ takes the same value
$j$, and thus $\Delta \tau_{j}$ is expressed as the correct sum of $\Delta t$
intervals.  With the help of (\ref{5.14}), (\ref{5.8}) can be written as
\begin{equation}
H_{0} = \frac{1}{2} \sum_{m=-\infty}^{\infty} \sum_{n=-\infty}^{\infty} g(I_{n}
- I_{m}) \Delta t_{m} \Delta t_{n}
\label{5.15}
\end{equation}
with sums over parabolas instead of particles. Similarly (\ref{2.10}) may be
written in the form
\begin{equation}
H_{1} = \sum_{m=- \infty}^{\infty} (I_{m} - I_{m-1}) h_{m},
\label{5.16}
\end{equation}
because $I_{m} - I_{m-1}$ is equal to the number of particles in parabola $m$.

\section{Contribution of $H_0$ to the Ground State Energy}
\label{sec4}

For an infinite configuration $\{ u_{j} \} _{j \epsilon {\Bbb Z}}$ of
particles,
expressions such as (\ref{2.1}) are formally infinite and do not possess a
direct interpretation. Following Aubry \cite{aubx}, we shall understand
(\ref{2.1}) as prescribing the change in energy if any finite set of particles
are displaced from one set of positions to another. A {\it {minimum energy}}
configuration is one for which any such change increases $H$ (or leaves it
constant) and a {\it {ground state}} is a minimum energy configuration which
satisfies certain recurrence conditions \cite{aubx}.  Aubry has shown
that under
suitable conditions (see also the additional remarks in App.~\ref{appa}) any
minimum energy configuration has a well-defined average separation between
particles,
\begin{equation}
\omega = Pw := \lim_{(k-j) \rightarrow \infty} (u_{k} - u_{j})/(k-j), 
\label{3.1}
\end{equation}
independent of how $j$ and $k$ behave individually (thus $j= -k$, or $j =k/2$
give the same result when $k \rightarrow \infty$). We shall call $w= \omega /
P$ the ``winding number", a notation clearly motivated by the twist map analogy
\cite{5}.

For a given $\omega$, there is a well-defined energy per particle
\begin{equation}
\epsilon = \lim_{(K-J) \rightarrow \infty} \frac{1}{(K-J)} \sum_{j=J}^{K-1}
[\frac{1}{2} (u_{j+1} - u_{j})^2 + V(u_{j})],
\label{3.2}
\end{equation}
where $\{ u_{j} \}$ is any minimum energy configuration for this $\omega$.
(Note that this is different from the process of choosing, for each $J$ and $K$
fixed, the $\{ u_{j} \}$ which minimizes the finite sum in (\ref{3.2}) ! )

If $w$ is irrational, any ground state configuration with $\omega = Pw$ has the
form
\begin{equation}
u_{j} = f_{\omega}(j \omega + c),
\label{3.3}
\end{equation}
where $c$ is some constant, and $f_{\omega}(z)$ is a monotone strictly
increasing ``hull" function \cite{aubx} which is step periodic, i.e.,
\begin{equation}
f_{\omega}(z+P) = P + f_{\omega}(z).
\label{3.4}
\end{equation}
When $f_{\omega}$ is discontinuous, as it is for the $V(u)$ considered here,
the discontinuities form a dense set, and there are two versions of
$f_{\omega}$: $f_{\omega}^{+}$, which at each discontinuity takes the maximum
possible value consistent with monotonicity, and is thus right-continuous, or
upper semi-continuous, and $f_{\omega}^{-}$, which takes the minimal possible
value at each discontinuity, and is thus left continuous, or lower
semi-continuous. Either $f_{\omega}^{+}$ or $f_{\omega}^{-}$ may be employed in
(\ref{3.3}), though not both simultaneously \cite{aubx}.

	When $w=p/q$ is rational, with $p$ and $q$ relatively prime integers
and $q>0$, one can again write ground state configurations in the form
(\ref{3.3}), where the function $f_{\omega}(z)$ is defined on the discrete set
of points $z=Ps/q$, where $s$ is any integer, and $c$ is an integer multiple of
$P/q$.   Defined in this way,
$f_{\omega}(z)$ is strictly increasing and satisfies (\ref{3.4}).  See
App.~\ref{appb} for the derivation of this result, which does not seem to be
explicitly stated in \cite{aubx}.  For convenience of exposition, we
shall assume
that $w$ is irrational in the following discussion, as the extension to the
case of rational $w$ is straightforward.

In a ground state configuration no $u_{j}$ can come closer than some fixed
finite distance to one of the dividing points $\bar{t}_{n}$, (\ref{2.6}) (see
App.~\ref{appa}).  Hence $f_{\omega}(z)$ will have jump discontinuities with
every $\bar{t}_{n}$ falling in the interior of one of these discontinuities, as
illustrated in Fig.~\ref{fig4}. As $f_{\omega}(z)$ is monotone, this means that
the $z$
axis is cut up into segments by dividing points
\begin{equation}
\bar{z}_{n} \leq \bar{z}_{n+1}; \quad \bar{z}_{n+N} = \bar{z}_{n} + P
\label{3.5}
\end{equation}
with the property that
\begin{equation}
z < \bar{z}_{n} < z' \ \Rightarrow \ f_{\omega}(z) < \bar{t}_{n} <
f_{\omega}(z').
\label{3.6}
\end{equation}
Thus there is a correspondence between intervals on the $z$ axis and those on
the $u$ axis in that $(\bar{z}_{n-1}, \bar{z}_{n})$ is mapped by $f_{\omega}$
into the interior of $(\bar{t}_{n-1}, \bar{t}_{n})$, so that intervals of type
$l$ on one axis correspond to intervals of type $l$ on the other.

	The preceding discussion allows us to find an explicit formula for the
integers $\{ I_{m} \}$ defined in (\ref{5.12}) in terms of the $\bar{z}_{m}$.
Assume the hull function is $f_{\omega}^{-}$. Then, see Fig.~\ref{fig4}, the
number
of the last particle falling in parabola $m$ (or preceding this parabola if it
is empty) is the largest integer $j$ for which
\begin{equation}
j \omega + c \leq \bar{z}_{m},
\label{5.17}
\end{equation}
which means
\begin{equation}
I_{m} = \hbox{Int}[(\bar{z}_{m} - c)/\omega].
\label{5.18}
\end{equation}
(If one employs $f_{\omega}^{+}$ rather than $f_{\omega}^{-}$, (\ref{5.17}) is
a strict inequality, and $\hbox{Int}$ in (\ref{5.18}) should be replaced by
$\hbox{int}$,
where $\hbox{int}(x)$ is the largest integer strictly less than $x$. The energy
per
particle is independent of the choice between $f_{\omega}^{+}$ and
$f_{\omega}^{-}$ .)
 
	The ground state energy per particle $\epsilon_{0}$ corresponding
to $H_{0}$, (\ref{5.15}), can be computed as follows: Consider a set of $LN$
successive parabolas, where $L$ is large. Suppose that two integers
$m_{0},n_{0}$ between $1$ and $N$ are given, and consider summands in
(\ref{5.15}) corresponding to pairs
\begin{equation}
(m,n) = (m_{0} + \mu N, n_{0} + (\mu + \nu) N),
\label{5.19}
\end{equation}
where $\mu$ and $\nu$ are integers, and $\mu$ is between $1$ and $L$. By
(\ref{2.4}), $\Delta t_{m}$ and $\Delta t_{n}$ are the same for all $m$ and $n$
of this type, while (\ref{3.5}), the step periodicity of the $\bar{z}_{n}$,
allows us to write
\begin{equation}
I_{n} - I_{m} = \hbox{Int}(D+A+ \mu/w) - \hbox{Int}(A +\mu/w ),
\label{5.20}
\end{equation}
where $w$ is $\omega / P$,
\begin{equation}
A = \zeta_{m_{0}}/w - c/\omega, \quad 
D = (\zeta_{n_{0}} - \zeta_{m_{0}} + \nu)/w,
\label{5.21}
\end{equation} 
and we have introduced the notation
\begin{equation}
\zeta_{n} := \bar{z}_{n}/P.
\label{5.23}
\end{equation}

As $\mu$ varies, the right side of (\ref{5.20}) takes on only two values:
$1+\hbox{Int}(D)$, which occurs for a fraction $\hbox{frac}(D)$ of the
values of
$\mu$,
assuming $w$ is irrational and $L$ is very large, and $\hbox{Int}(D)$, which
occurs
for a fraction $1-\hbox{frac}(D)$ of the $\mu$ values. Hence the sum of
$g(I_{n}
-
I_{m})$ over the pairs (\ref{5.19}) with $\nu$ fixed is for large $L$ given by
$Lg(D)$, using the extended definition of $g$ in (\ref{5.11}).

What remains is a sum over $\nu$, which can be extended from $-\infty$ to
$\infty$ with negligible error because $g$ cuts off exponentially, and sums
over $m_{0}$ and $n_{0}$. Upon dividing by the numbers of particles $LP /
\omega = L / w$ in a system of length $LP$, one obtains
\begin{equation}
\epsilon_{0} = \frac{1}{2} \sum_{m=1}^{N} \sum_{n=1}^{N}
\Delta t_{m} \Delta t_{n} {\cal G} (\zeta_{n} - \zeta_{m})  
\label{5.24}
\end{equation}
for the energy per particle where  the subscripts of (\ref{5.19})
have been omitted from $m$ and $n$, and  we have introduced a new function
\begin{equation}
{\cal G} (x) = w \sum_{\nu = - \infty}^{\infty} g[\frac{x+\nu}{w}]
\label{5.25}
\end{equation}
which depends implicitly on both $w$ and $\kappa$ (or $r$). Note that
\begin{equation}
{\cal G} (x) = {\cal G} (1+x) = {\cal G} (-x)
\label{5.26}
\end{equation}
is periodic and symmetrical about $x=0$ and $x=0.5$, and convex for $x$ between
$0$ and $1$, as depicted in Fig.~\ref{fig6}.

\section{Finding the Ground State}
\label{sec5}
The energy per particle $\epsilon_{1}$ corresponding to $H_{1}$, (\ref{2.10}),
is easily evaluated when $w$ is irrational by noting that the sequence of
equally spaced points on the $z$ axis which form the arguments of $f_{\omega}$
in (\ref{3.3}) will eventually approach a uniform distribution if mapped modulo
$P$ into the interval $[0,P)$. That is to say, the fraction of these points
falling inside an interval of type $l$ on the $z$ axis, which is  the same
as the fraction of particles whose positions lie in parabolas of type
$l$, is equal to
\begin{equation}
\psi_{l} = \zeta_{l} -\zeta_{l-1},
\label{6.1}
\end{equation}
with $\zeta_{l}$, (\ref{5.23}), equal to $\bar{z}_{l} / P$. 
In view of (\ref{3.5}),
\begin{equation}
\zeta_{n+N} = 1+ \zeta_{n}
\label{6.5}
\end{equation}
and therefore
\begin{equation}
\sum_{l=1}^{N} \psi_{l} = 1.
\label{6.6}
\end{equation}
As a consequence,
\begin{equation}
\epsilon_{1} = \sum_{l=1}^{N} h_{l} \psi_{l} =
- \sum_{l=1}^{N} \eta_{l} \zeta_{l} +h_{1}
\label{6.2}
\end{equation}
where
\begin{equation}
\eta_{l} = h_{l+1} - h_{l},
\label{6.3}
\end{equation}
and because the $h$'s are periodic, (\ref{2.5}),
\begin{equation}
\sum_{l=1}^{N} \eta_{l} = 0.
\label{6.4}
\end{equation}
  
Thus the total energy per particle $\epsilon = \epsilon_{0} + \epsilon_{1}$ can
be written in the form
\begin{equation}
\epsilon = \frac{1}{2} \sum_{n=1}^{N} (\Delta t_{n})^2 {\cal G}(0) +
\sum_{m=1}^{N-1} \sum_{n=m+1}^{N} \Delta t_{m} \Delta t_{n}
{\cal G} (\zeta_{n} - \zeta_{m}) - \sum_{m=1}^{N} \eta_{m} \zeta_{m} + h_{1}
\label{6.7}
\end{equation}
by combining (\ref{5.24}), slightly rewritten using (\ref{5.26}), with
(\ref{6.2}).  Here $h_1$ is an additive constant which is set equal to zero in
the following discussion.  Then $\epsilon$ is a function of the unknown
parameters $\zeta_{1},
\zeta_{2}, ... , \zeta_{N}$, and must be minimized by varying these parameters.
In fact, adding the same constant to every $\zeta_n$ leaves $\epsilon$
unchanged (note Eq. \ref{6.4}) and therefore, since the $\psi$'s, (\ref{6.1}),
cannot be negative, it suffices to consider the problem of minimizing
$\epsilon$ on the{\it { simplex}}
\begin{equation}
0= \zeta_{0} \leq \zeta_{1} \leq \zeta_{2} \leq \ ... \ \leq
\zeta_{N-1} \leq \zeta_{N} = 1.
\label{6.8}
\end{equation}
 
One can think of the $\{ \zeta_{m} \}$ as the positions of a set of $N$
``quasi-particles" located on a circle of unit circumference, interacting with
each other through a set of pair potentials given by ${\cal G}$ and, in
addition, subjected to constant single-particle forces (the $\eta$'s), whose
sum is zero. The fact that ${\cal G}(x)$ has a minimum at $x = \frac{1}{2}$
means that the pair forces tend to keep the quasi-particles separated, although
the imposition of suitable single-particle forces can push two or more
together. However, the order (\ref{6.8}) must be preserved: these
quasi-particles have hard cores and will not move through each other.
  
	Note that $\epsilon$, regarded as a function of the $\zeta_{j}$ on the
domain (\ref{6.8}), is a sum of continuous convex functions, and is therefore
continuous and convex. As a continuous function on a compact domain, it
necessarily has a minimum someplace on (\ref{6.8}), possibly on the boundary.
Either $\epsilon$ takes its minimum at a unique point or on some larger convex
subset of (\ref{6.8}). If $w$ is a rational number one can find examples where
the minimum is not unique. These correspond to (first-order) phase transitions
where one can have two or more distinct ground state configurations. On the
other hand, if $w$ is irrational, $\epsilon$ is a strictly convex function
which achieves its minimum at a unique point in the simplex (\ref{6.8}). The
strict convexity of $\epsilon$ in this case arises from the strict convexity of
${\cal G}(x)$, whose derivative with respect to $x$ has a dense set of
discontinuities on the interval $(0,1)$ when $w$ is irrational.

	Once the $\{ \zeta_{m} \}$ are known, and thus the $\{\bar z_m\}$, see
(\ref{5.23}), the ground state configuration $\{u_j\}$ and the hull function
$f_{\omega}$, (\ref{3.3}) can be obtained by means of the following
construction.  Define the piecewise constant and step periodic function
\begin{equation}
	T(z) = t_{m} \hbox{ for }\bar{z}_{m-1} < z \leq \bar{z}_{m},
\label{7.1}
\end{equation}
see Fig.~\ref{fig7}, so that in the ground state
\begin{equation}
\tau_{k} = T(k \omega + c).
\label{7.2}
\end{equation}
Inserting this expression in (\ref{5.7}) yields
\begin{equation}
u_{j} = \sum_{k= - \infty}^{\infty} g(j-k) T(k \omega + c) =
\sum_{l= - \infty}^{\infty} g(l) T(j \omega + c - l \omega).
\label{7.3}
\end{equation}
Comparison with (\ref{3.3}) shows that
\begin{equation}
f_{\omega}^{-}(z) = \sum_{l= - \infty}^{\infty} g(l) T(z-l \omega),
\label{7.4}
\end{equation}
where the sum yields $f_{\omega}^{-}$ because $T$ has been defined to be
continuous from the left. If at each discontinuity one sets $T(\bar{z}_{m})$
equal to $t_{m+1}$ rather than $t_{m}$, (\ref{7.4}) will yield
$f_{\omega}^{+}$.

\section{Conclusion}
\label{concl}  
	We have shown how the problem of finding ground state configurations
for a Frenkel-Kontorova model in a system of parabolas of identical curvature,
assuming the system has a finite period, can be reduced to finding the minimum
of a certain convex function, (\ref{6.7}), over the variables $ \{ \zeta_{n}
\}$ belonging to a finite simplex (\ref{6.8}). While this second minimization
problem is not trivial, the examples in our previous paper \cite{14} show that
the solution can be worked out in particular cases with a certain amount of
effort, for instance, by treating the $\{ \zeta_{n} \}$ as parameters and
computing the corresponding $\{ \eta_{n} \}$. An efficient numerical procedure
for finding the minimum, which works well for $N$ up to the order of 200, is
discussed in the paper which follows \cite{SUGS96}.

	It is natural to ask whether further generalizations of this type of
model might prove interesting. There are several possibilities worth
considering: First, one can imagine taking the limit $N \rightarrow \infty$
with $P$ held finite in such a way that $V$ converges to some {\it {smooth or
piecewise smooth}} periodic potential, such as a cosine. While it is unlikely
that such an approach will yield an exact solution for the ground state of an
arbitrary smooth periodic potential, it may nonetheless provide some
interesting insights, and preliminary studies of this problem are
encouraging, see \cite{sch91}.

	Second, one can imagine a limit with both $N$ and $P$ tending to
infinity in a way which leaves their ratio fixed, thus yielding an example of a
potential, still composed of parabolas, which is {\it {quasiperiodic}} or
possesses some other deterministic form of non-periodicity. An exact solution
for such a case would be quite interesting.

	Third, one might hope to study cases in which $V$ consists of parabolas
of two (or more) different curvatures; for example, parabolas which meet at
points where their first derivatives are continuous and their second
derivatives discontinuous. Such models have been studied numerically
\cite{19a}. An interesting analytical investigation of a particular case and
the corresponding ``tent map" has been performed by Bullett \cite{20}.  The
basic difficulty with non-uniform curvature is that the Green's function needed
to solve the counterpart of (\ref{5.1}) is much more difficult to calculate.
Yet the fact that novel phenomena appear already when the basic $N=1$ parabolic
model is extended to $N=2$ suggests that any progress in this direction would
be worth the effort.

\acknowledgments

We are grateful to S.Aubry for bringing Ref.~\cite{16} to our attention.  One
of us (R.B.G.) would like to thank the Institut de Hautes Etudes Scientifiques
in Bures-sur-Yvette and the Service de Physique Th\'eorique of the C.E.N.
Saclay for their hospitality during a sabbatical year, and acknowledge
financial support by the U. S. National Science Foundation through grant
DMR-9009474.

\appendix
\section{Minimum Distance of Approach to Dividing Points.}
\label{appa}
                                                     
Aubry's theorems \cite{aubx} for minimum energy and ground state configurations
require that $V(u)$ be a smooth function, a condition violated at dividing
points, (\ref{2.6}), where the first derivative of the potential is
discontinuous. This difficulty is easily circumvented by noting that, as
demonstrated below, no particle in a minimum energy configuration can come
closer than a minimal distance $\delta_{0} > 0$ to a dividing point.
Consequently one can always suppose, in order to apply Aubry's arguments, that
$V(u)$ has been replaced by a smooth $\bar{V}(u)$ identical to $V(u)$ except
for $u$ closer to a dividing point than $\delta_{0}$, with $\bar{V}$ always
greater than or equal to $V$ (see Fig.~\ref{fig8}). Minimum energy
configurations of
$\bar{V}$ are identical to those of $V$, as otherwise one could lower the
energy by moving a single particle, following the reasoning given below.

Assume that a dividing point occurs at $u=0$, and that particle $1$ of a
minimum energy configuration falls in the parabola to the right of this point,
at $u_{1} = b > 0$. Ignoring an additive constant, we can write, for $u$ near
zero,
\begin{equation}
	V(u) = \cases{ \kappa u^2/2 - \alpha u & for $ u > 0,$ \cr
		\kappa u^2/2 + ( \gamma - \alpha )& for $u < 0, $ \cr}
\label{A.1}
\end{equation} 
 with $\alpha$ a constant and $\gamma > 0$ the discontinuity
in $V'(u)$ at $u=0$.

If particles $u_{0}$ and $u_{2}$ are held fixed, the part of the energy $H$,
(\ref{2.1}), involving $u_{1}$ can be written as $\hat{V}(u_{1})$, where,
omitting an additive constant,
\begin{equation}
\hat{V} (u) = V(u) + u^2 - ( u_{0} + u_{2}) u .
\label{A.2}
\end{equation}
Of course $\hat{V}(u)$ must be a minimum at
\begin{equation}
u_{1} = b = (u_{0} + u_{2} + \alpha)/(2 + \kappa) > 0,
\label{A.3}
\end{equation}
where it takes the value
\begin{equation}
- (u_{0} + u_{2} + \alpha))^2/(4+2 \kappa).
\label{A.4}
\end{equation}
If $b$ is smaller than $\gamma / (2+\kappa)$, $\hat{V}(u)$ also has a local
minimum at
\begin{equation}
u_{1}' = (u_{0} + u_{2} + \alpha - \gamma)/(2 + \kappa) < 0,
\label{A.5}
\end{equation}
where it takes the value
\begin{equation}
- (u_{0} + u_{2} + \alpha - \gamma)^2/(4+2 \kappa).
\label{A.6}
\end{equation}

Now (\ref{A.6}) cannot be less than (\ref{A.4}), as otherwise the energy would
be lowered by changing $u_{1}$ to $u_{1}'$.  (The argument is not affected by
the possible presence of another dividing point between $u_{0}$ and $u_{1}'$,
as in that case the energy at $u_{1}'$ would be even less than (\ref{A.6}).)
Consequently one has
\begin{equation}
|u_{0} + u_{2} + \alpha - \gamma | = | (2 + \kappa )b - \gamma | \leq u_{0} +
u_{2} + \alpha = (2 + \kappa ) b,
\label{A.7}
\end{equation}
and thus a lower bound
\begin{equation}
\gamma \leq 2(2+ \kappa ) b 
\label{A.8}
\end{equation}
for $b$. As the discontinuity $\gamma$ in $V'(u)$ at the dividing point
$\bar{t}_{j}$ is $\kappa \Delta t_{j}$, we conclude that no particle in a
minimum energy configuration can come closer than
\begin{equation}
\delta_{0} = [ \frac{\kappa}{4 + 2 \kappa} ] \ \min_{j \epsilon {\Bbb Z}} 
\Delta t_{j}
\label{A.9}
\end{equation}
to one of the dividing points.

\section{Hull Function for Rational Winding Number}
\label{appb}

	It is sufficient to consider the case in which $P=1$ and $c=0$, and as
$w=p/q$ is fixed, the subscript $\omega$ can be omitted from the hull function
$f$.  As shown in \cite{aubx}, a ground state configuration $\{ u_j\}$ is
periodic, with
\begin{equation}
	u_{j+q} = u_j + p
\label{B.1}
\end{equation} 
for every $j$.  Therefore, for any integer $m$, we can define
\begin{equation}
	f(m/q) = u_j - k,
\label{B.2}
\end{equation}
where $j$ and $k$ are any two integers satisfying
\begin{equation}
	jp-kq=m.
\label{B.3}
\end{equation}
The fact that $p$ and $q$ are relatively prime means that (\ref{B.3}) always
has solutions, while (\ref{B.1}) ensures that two solutions yield the same
result when inserted in (\ref{B.2}), and that (\ref{3.4}) is satisfied with
$P=1$: 
\begin{equation}
	f(1+z) = 1+ f(z). 
\label{B.4}
\end{equation}

	Let $s$ and $t$ be two integers satisfying 
\begin{equation}
	sp-tq = 1,
\label{B.5}
\end{equation}
and define a second ground state configuration $\{ v_j\}$ by means of:
\begin{equation}
	v_j = u_{j+s} - t.
\label{B.6}
\end{equation}
Using (\ref{B.2}), (\ref{B.5}), and (\ref{B.6}) one finds that:
\begin{equation}
	u_{ms} = f({m\over q}) + mt,\quad v_{ms} = f({m+1\over q}) +mt.
\label{B.7}
\end{equation}
A consequence of the property of ``total ordering'' of ground states which is
proved in \cite{aubx} is that if $v_j\leq u_j$ for one value of $j$, this must
also hold for all values of $j$. Applied to (\ref{B.7}), this means that if
$f[(m+1)/q]$ is less than or equal to $f(m/q)$ for one value of $m$, the same
is true for all values of $m$, a result which clearly contradicts (\ref{B.4}).
Hence
\begin{equation}
	f({m+1\over q}) > f({m\over q})
\label{B.}
\end{equation}
for every $m$, and $f$ is a strictly increasing function on the discrete
set where it is defined.

\begin{figure} 
\caption{
 Labeling of the minima and dividing points for successive 
parabolas forming $V(u)$.}
\label{fig3}
\end{figure}

\begin{figure} 
\caption{
 The function $g(x)$ for $\kappa = 0.5$.}   
\label{fig5}
\end{figure}

\begin{figure} 
\caption{
 A hull function $f_{\omega} (z)$ showing the 
relationship of the dividing points $\bar{z}_{n}$ and $\bar{t}_{n}$.}
\label{fig4}
\end{figure}

\begin{figure} 
\caption{
 The function ${\cal G}(x)$ for $\kappa = 0.2$, $w$ equal
to the golden mean (0.618...).}
\label{fig6}
\end{figure}

\begin{figure} 
\caption{
 The function $T(z)$, see text.}   
\label{fig7}
\end{figure}

\begin{figure} 
\caption{
The smooth potential $\bar{V}(u)$, dashed curve, is identical to
$V(u)$, solid curve, except near the dividing points of the latter.}
\label{fig8}
\end{figure}

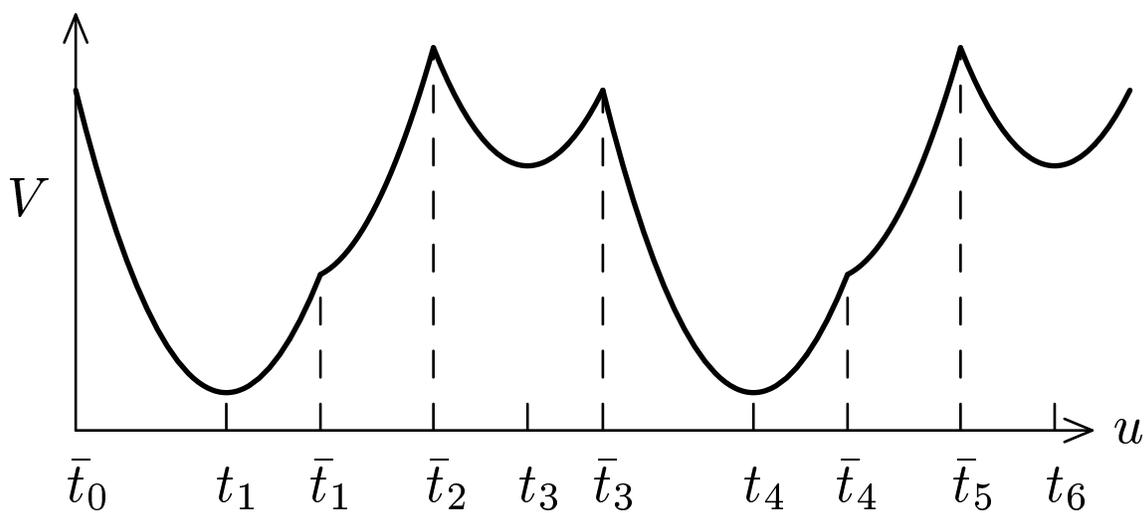
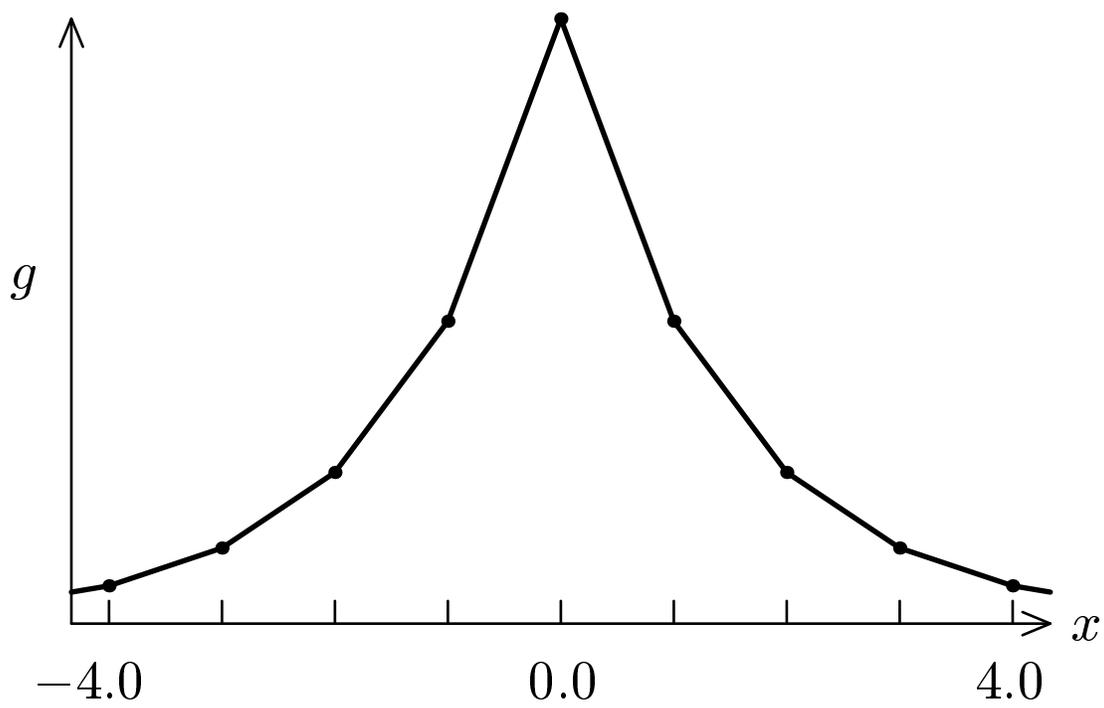
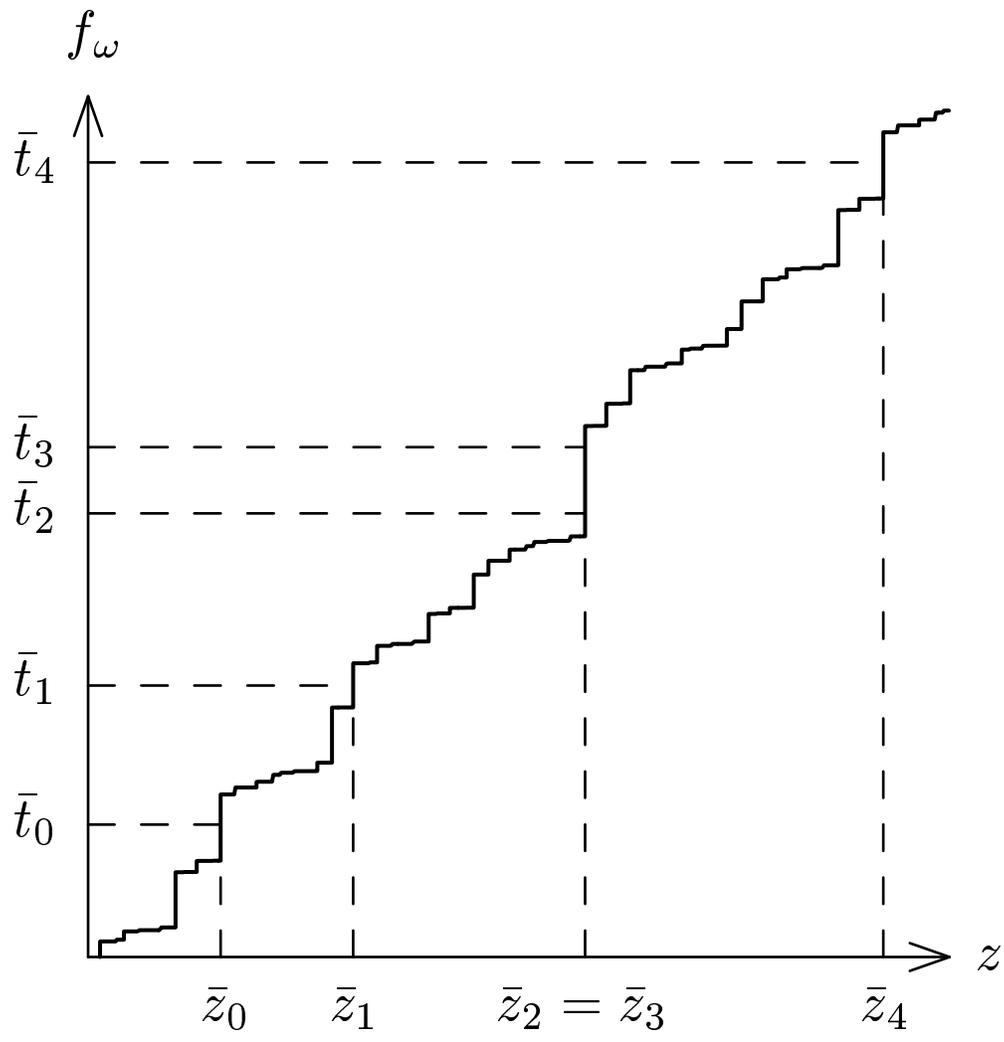
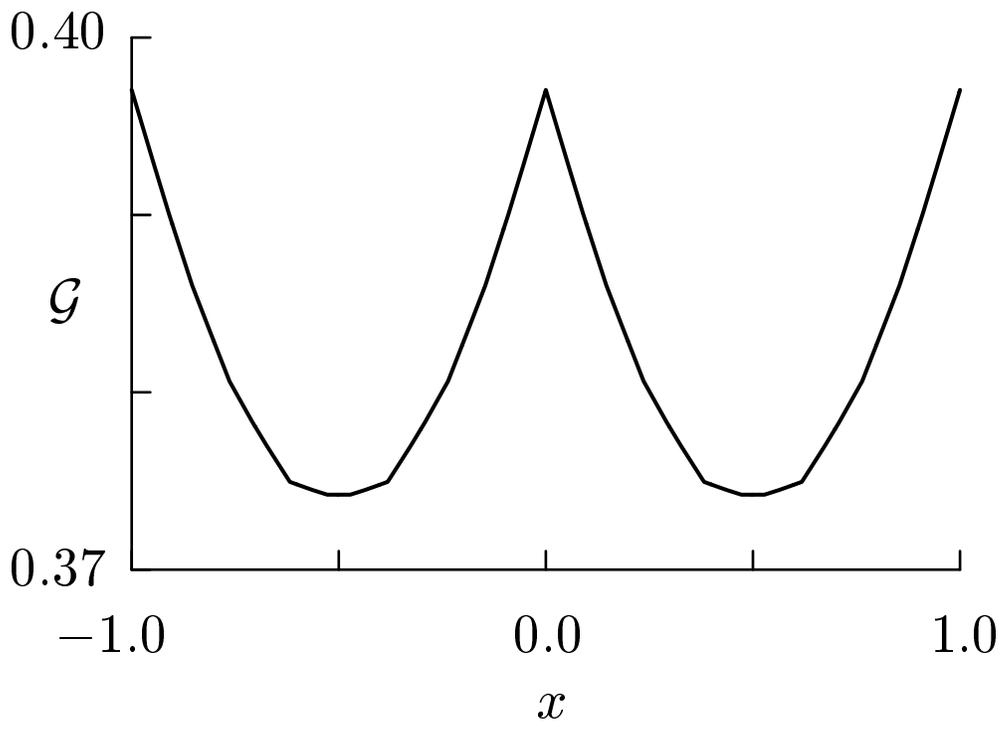
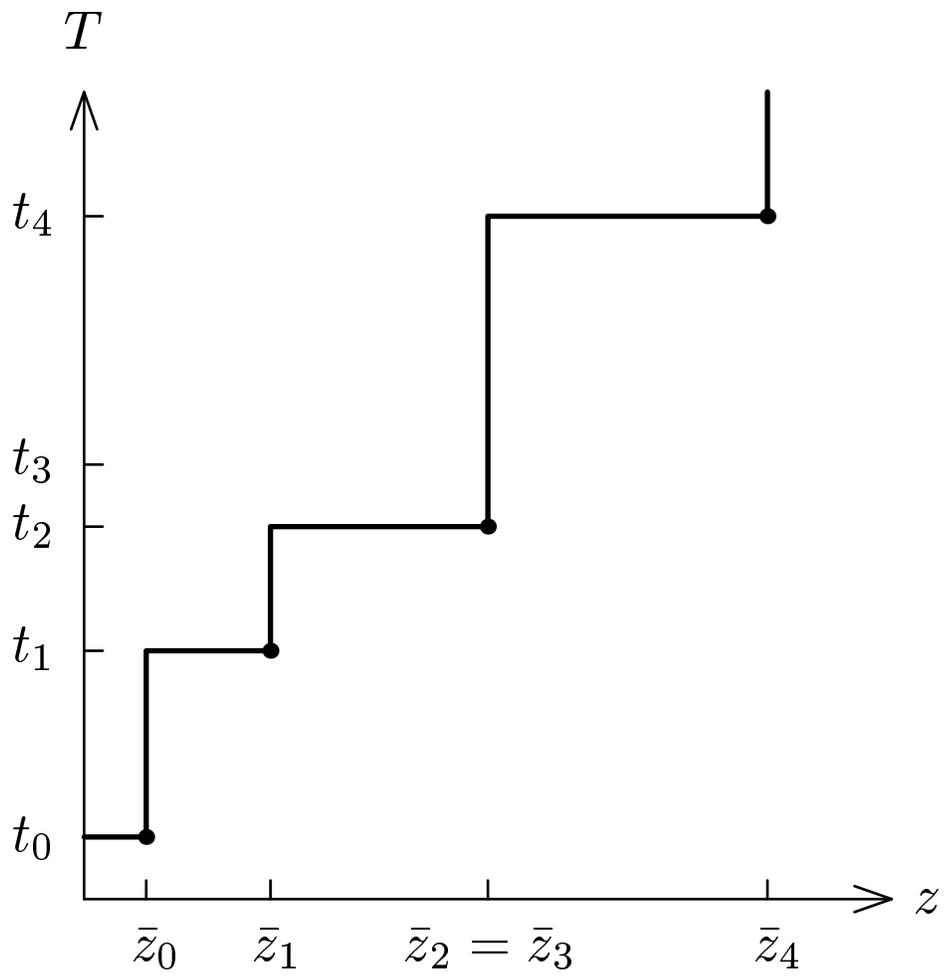
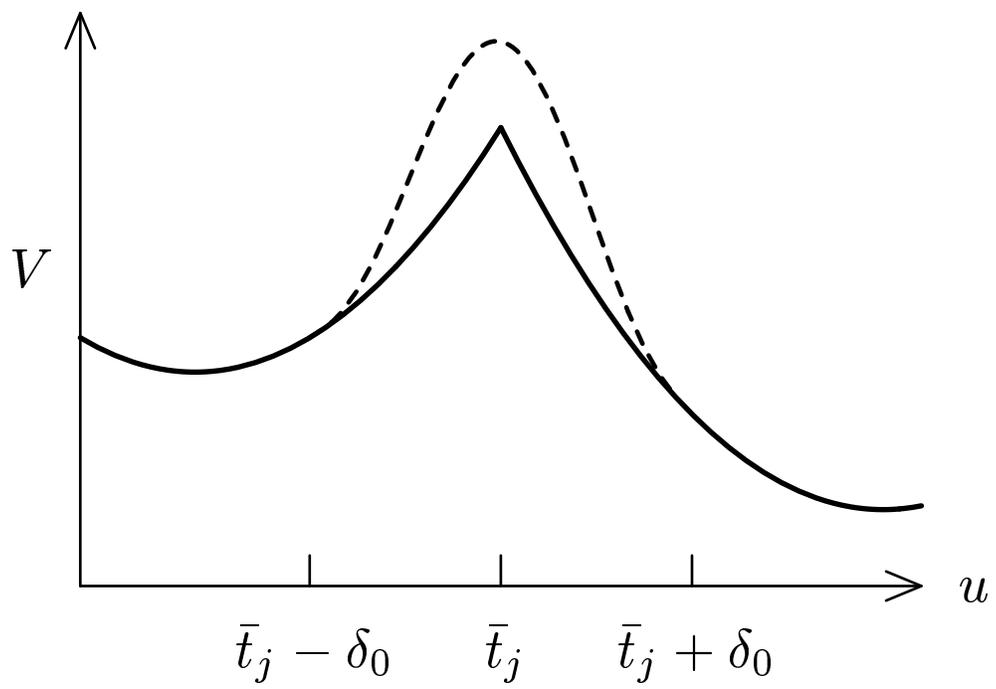
\end{document}